\documentclass[11pt,a4paper]{emulateapj}
\usepackage{natbib}
\input{psfig.sty}
\singlespace
\newcommand{\kms}{\,{\rm km\,s^{-1}}}
\newcommand{\msun}{\,{\rm M_\odot}}

\newcommand{\etal}{{et al.\ }}

\newcommand{\gas}{{\rm gas}}

\newcommand{\beq}{\begin{equation}}
\newcommand{\eeq}{\end{equation}}
\newcommand{\ba}{\begin{eqnarray}}
\newcommand{\ea}{\end{eqnarray}}
\def\spose#1{\hbox to 0pt{#1\hss}}
\newcommand{\lta}{\mathrel{\spose{\lower 3pt\hbox{$\mathchar"218$}}
      \raise 2.0pt\hbox{$\mathchar"13C$}}}
\newcommand{\gta}{\mathrel{\spose{\lower 3pt\hbox{$\mathchar"218$}}
      \raise 2.0pt\hbox{$\mathchar"13E$}}}
\lefthead{Volonteri \& Rees}
\righthead{Luminous quasars at z=6}
\makeatletter

\newenvironment{figurehere}
  {\def\@captype{figure}}
  {}
\makeatother
\begin{document}
\title{Quasars at z=6: the survival of the fittest}
\author{Marta Volonteri\altaffilmark{1} \& Martin J. Rees\altaffilmark{1}}
\altaffiltext{1}{Institute of Astronomy, Madingley Road, Cambridge CB3 0HA, UK}

\received{---------------}
\accepted{---------------}

\begin{abstract}
The Sloan Digital Sky survey detected luminous quasars at very high redshift, $z>6$. Follow-up observations
indicated that at least some of these quasars are powered by supermassive black holes (SMBHs) with masses in
excess of $10^9\msun$. SMBHs, therefore, seem to have already existed when the Universe was less than {\rm 1 Gyr} old, and the bulk of galaxy formation still has to take place. We investigate in this paper to which extent accretion and dynamical processes influence  the early growth of SMBHs. We assess the impact of (i) black hole mergers, (ii) the influence of the merging efficiency and (iii) the negative contribution due to dynamical effects which can kick black holes out of their host halos (gravitational recoil). We find that if accretion is always limited by the Eddington rate via a thin disc, the maximum radiative efficiency (spin) allowed to reproduce the LF at $z=6$ is $\epsilon=0.12$ ($\hat a=0.8$), when the adverse effect of the gravitational recoil are taken into consideration. Dynamical effect unquestionably  cannot be neglected in studies of high-redshift SMBHs.  If black holes can accrete at super-critical rate during an early phase, reproducing the observed SMBH mass values  is not an issue, even in the case that the recoil velocity is in the upper limits range, as the mass ratios of merging binaries are skewed towards low values, where the gravitational recoil effect is very mild. We propose that SMBH growth at early times is very selective, and efficient only for black holes hosted in high density peak halos. 

\end{abstract}

\section{Introduction}
It seems a challenge for theoretical models to explain the luminosity function
of luminous quasars at $z\approx 6$ in the Sloan Digital Sky Survey \citep[SDSS][]{Fanetal2001a}. 
The luminosities of these quasars, well in excess of $10^{47}$ erg/s, imply SMBHs with 
masses $~10^9 M_\odot$ already in place when the Universe is only {\rm 1 Gyr} old. 
The accretion of mass at the Eddington rate causes the black hole (BH) mass to increase 
in time as
\beq
M(t)=M(0)\,\exp\left(\frac{1-\epsilon}{\epsilon}\frac{t}{t_{\rm Edd}}\right),
\eeq
where $t_{\rm Edd}=0.45\,{\rm Gyr}$ and $\epsilon$ is the radiative
efficiency. Among the seed BHs proposed, the more commonly invoked (e.g. PopIII star remnants, 
gravitationally collapsed star clusters) have masses in the range $M(0)=10^2-10^4 M_\odot$, 
forming at $z=30$ or less. Given $M(0)$, the higher the efficiency, the longer it takes 
for the BH to grow in mass by (say) 20 e-foldings \citep{Shapiro2005}. For a Schwarzschild black hole,
the standard thin disc radiative efficiency is $\epsilon\approx0.06$, and there is
plenty of time for the BH seed to become supermassive. The timescale to grow from $M(0)=10^2-10^4 M_\odot$
to $\simeq 10^9 M_\odot$ is less than  {\rm 0.5 Gyr}. 

If accretion is via a geometrically thin disc, though, the alignment of a SMBH with the angular 
momentum of the accretion disc tends to efficiently spin holes up (Volonteri et al 2005), 
and radiative efficiencies can therefore approach 30-40\%. With such a high efficiency, $\epsilon=0.3$, 
it takes longer than  {\rm 2 Gyr} for the seed to grow up to a billion solar masses.

It seems therefore difficult to reproduce the observational constraints without invoking exotic 
processes.

We take here a conservative approach and try to determine the parameter space
which allows a SMBH growth compatible with observational constraints at $z=6$,
as evinced from the luminosity function of quasars \citep{Fanetal2004} and 
observations of galaxies at $z\simeq 6$.
In this paper we critically assess models for the early evolution of SMBHS, exploring
the parameter space of the involved processes: accretion rate, radiative efficiency,
dynamical processes and the initial density of massive black hole (MBH) seeds. 

%The first variables at play are the mass and frequency of MBH seeds. An alternative to 
%PopIII remnants are seed holes formed via direct collapse of dense gas in the center of 
%low angular momentum halos. Though this is a very interesting and promising alternative, 
%it is not clear yet how effective the transport of angular momentum is, at the center of 
%these halos. 

At $z<5$ MBH mergers do not play a fundamental role in building up the mass of SMBHs
(Yu \& Tremaine 2003), but they can be possibly important at $z>5$, where we do not 
have constraints from a Soltan-type argument,  which compares the
local MBH mass density with the mass density inferred from luminous
quasars, as the luminosity function of quasars is not constrained at $z>6$.
Mergers can possibly contribute positively to the build-up of the high redshift 
SMBH population (Yoo \& Miralda-Escud\'e 2004), as they contribute to make a big black hole 
from many small seeds. 

On the other hand, dynamical processes can disturb the 
growth of BHs, especially at high redshift (Haiman 2004, Yoo \& Miralda-Escud\'e 2004), 
and give a negative contribution to SMBH growth.
In the shallow potential wells of mini-halos, the growth of MBHs can be halted by the
`gravitational rocket' effect, the recoil due to the non-zero net linear momentum 
carried away by gravitational waves in the coalescence of two black holes. 
Also, if MBHs are widespread, and binary black holes coalescence timescales are long enough 
for a third MBH to fall in and interact with the central binary, Newtonian three-body 
interactions can lead to the expulsion, or recoil, of the binary. The accretion history must 
then be studied jointly with the dynamics involving MBHs.

Yoo \& Miralda-Escud\'e (2004) explored the minimum conditions that would allow the growth
of seed MBHs up to the limits imposed by the highest redshift quasar currently known: 
SDSS 1148+3251. This quasar, at $z=6.4$, has estimates of the SMBH mass in the range 
$(2-6)\times 10^9 M_\odot$ \citep{Barthetal2003, Willottetal2003}. 
Yoo \& Miralda-Escud\'e (2004) showed that the mass of this SMBH can be explained
assuming (i) continued Eddington-limited accretion onto MBHs forming in halos with
$T_{\rm vir} >2000$K at $z\leq 40$. Their model assumes, also, (ii)  a 
high influence of BH mergers in increasing the MBH mass: a contribution by itself 
of order $10^9\msun$. Their investigation takes into account the negative feedback
due to  the `gravitational rocket' effect (see also Haiman 2004).
Recent estimates suggest a more modest recoil velocity, compared to the typical values 
adopted in the past. We test here how much difference (if any) these new estimates imply 
for the growth of black holes in pre-galactic halos.

It is also important to understand where $z=6$ quasars are hosted. By matching the number 
density in haloes more massive than $M_h$ to the space density of quasars at $z=6$ 
\citep{Fanetal2004}, $M_h=10^{13}\msun$. This assumption corresponds to requiring 
that the duty cycle of high redshift quasars is unity, i.e. all the BHs inhabiting 
halos with mass larger than $10^{13}\msun$ are active. The only available observation of 
a quasar host \citep{Walteretal2004}, shows interesting features. The kinematics of the observed 
molecular gas implies the lack of a massive bulge around the SMBH, but suggests a DM halo 
with a mass similar to that of the predicted bulge, of order $10^{12}\msun$. This is much 
less than $M_h=10^{13}\msun$ and allows for a much smaller duty cycle. The uncertainties 
on the dynamical configuration of the gas (e.g. inclination) are still large, however.

In the next section we review the model for assembly of MBHs in cold dark matter (CDM)
cosmogonies.  We then review models for MBH growth by accretion in high redshift halos 
(\S3), and discuss the dynamical evolution of MBH binaries in pre-galactic systems (\S4).
The results from the interplay of accretion and dynamical processes are presented in 
\S5. Finally, we discuss the implications on the global evolution of the SMBH population (\S6). 
All results shown below refer to a $\Lambda$CDM world model with $\Omega_M=0.3$, $\Omega_\Lambda=0.7$, 
$h=0.7$, $\Omega_b=0.045$, $\sigma_8=0.93$, and $n=1$.

\section{Assembly of pregalactic MBH\lowercase{s}}

The main features of a plausible scenario for the hierarchical
assembly, growth, and dynamics of MBHs in a $\Lambda$CDM cosmology
have been discussed in \cite{VHM,Volonterietal2005,Madauetal2004}.  Dark matter halos and their
associated galaxies undergo many mergers as mass is assembled from
high redshift to the present.
The halo merger history is tracked
backwards in time with a Monte Carlo algorithm based on the extended
Press-Schechter formalism. ``Seed" holes are assumed to form with intermediate masses
in the rare high $\nu-\sigma$ peaks  collapsing at $z=20-25$
\citep{MadauRees2001} as end-product of the very first generation of stars.

%Pregalactic seed IMBHs form within the mass ranges $20<m_\bullet<70\,\msun$
%and $130<m_\bullet<600\,\msun$ (Fryer, Woosley \& Heger 2002, Omukai \& Palla 2003), 
%as remnants of the first generation of massive metal-free stars that do not disappear 
%as pair-instability supernovae.   
%For simplicity, within these intervals the differential black hole mass 
%function is assumed to be flat. 

%i.e. the IMF-averaged BH mass is
%335$\,\msun$. 

%The first stars and the first black holes form within mini-halos 
%above the cosmological Jeans mass collapsing at $z>20$ 
%from rare $\nu$-$\sigma$ peaks of the primordial density field. 
As our fiducial model we take $\nu=4$ at $z=24$, corresponding to a mass density parameter in
MBHs of order $\rho_\bullet\simeq 10^{2}\msun Mpc^{-3}$, about three orders of magnitude smaller than the local SMBH mass 
density \citep{Aller2002, Merlonietal2004, Marconietal2004, YuTremaine2002}. The mass density in seed MBHs cannot be 
larger than the mass density of SMBHs at $z=3$, $\rho_{BH}(z=3)\simeq 4.5\times 10^{4}\msun Mpc^{-3}$
\citep{Merloni2004}. 
%From z=3 to z=0, SMBHs accrete, in quasar phase, about $1.5-4.5\times 10^{5}\msun Mpc^{-3}$.  
%\beq
%\Omega_\bullet={0.00047\,\Omega_M\,\langle m_\bullet\rangle\over 
%M_{\rm seed}}~10^{-9}. \label{ombh}
%\eeq
%This is much less than the mass density of the supermassive variety 
%found in the nuclei of most nearby galaxies, $\Omega_{\rm SMBH}=(2.1\pm
%0.3) \times 10^{-6}$ (Yu \& Tremaine 2002).  
%Note that 

This choice of seed MBHs occupation is similar to that of \cite{VHM}
(seed holes in $3.5\sigma$ peaks at $z=20$), ensuring that galaxies hosted in halos with mass 
larger than $10^{11}\msun$ are seeded with a MBH. The assumed threshold allows efficient formation 
of SMBHs in the range of halo masses effectively probed by dynamical studies of SMBH hosts 
in the local universe. 
We will also consider a case in which seed holes are more numerous (e.g. 25 times more than 
the fiducial case). 

We are here interested in the evolution of the uttermost massive halos present at $z=6$, 
i.e. very highly biased structures \citep{Diemandetal2005}. As a consequence,  the density of seed 
holes within the volume occupied by the progenitors of  the halo is large enough that the merging 
of two minihalos both hosting a BH is not a rare event. This is in contrast with the 
\emph{average} density of seed holes. In a cosmic volume, the seed holes typically evolve in 
isolation \citep[cfr.][]{Madauetal2004}.

We generate Monte Carlo realizations (based on the extended Press-Schechter formalism) of the 
merger hierarchy of a $M_h=10^{13}\msun$ halo at $z=6$. The halo mass is chosen by requiring that 
the number density in haloes more massive than $M_h$ matches the space density of quasars at $z=6$ 
\citep{Fanetal2004}.

\section{Accretion and radiative efficiency}
We explore here two regimes: in one case the accretion rate is limited to the 
Eddington rate, in the second case MBHs are allowed to accrete at super-critical 
rates (cfr. \cite{VolonteriRees2005} for a detailed description of the model). 
In all cases we assume that accretion episodes are triggered by major mergers \citep{MihosHernquist1994, 
MihosHernquist1996, DiMatteo2005, Steliosetal2005}, which we define as mergers between halos with a mass ratio 1:10 or higher. Mergers with smaller mass ratios are probably unable to trigger substantial gas inflow \citep{tjphd}.

In the first case, the hole accretes at the Eddington rate a gas rest 
mass $\Delta m_0$. This leads to a change in the total mass-energy of the hole given by
\begin{equation}
\Delta m=7.7\times 10^5\,\msun~V_{c,100}^{4.6},
\label{macc_eq}
\end{equation}
where $V_{c,100}$ is the circular velocity of the merged system in
units of 100 $\kms$. Adopting Equation \ref{macc_eq} implies assuming that the 
correlations between black hole mass, velocity dispersion and circular velocity 
are maintained throughout cosmic time \citep{Tremaineetal2002, Ferrarese2002}. 
The quantities $\Delta m$ and $\Delta m_0$ are related by 
$\Delta m=(1-\epsilon)\,\Delta m_0$, where $\epsilon$ is the mass-to-energy 
conversion efficiency, equal for thin-disk accretion to the binding energy per 
unit mass of a particle in the last stable circular orbit\footnote{The simple relation 
would be modified when the thickness of the disc is of order its radius 
and can also changed by magnetic effects which allow energy release 
from within the innermost stable orbit \citep[e.g.,][]{Kroliketal2005}}. The MBHs spin, $S$, is modified 
during the accretion phase as described in \cite{Volonterietal2005}. We are interested 
here in determining the maximum radiative efficiency (or spin, using the standard conversion) 
which allows the growth of MBHs into SMBHs with masses larger than $10^9\msun$ 
before $z=6$. As a reference, we have considered an upper limit to the 
radiative efficiency of $\epsilon=0.16$. This corresponds, adopting the standard conversion 
for accretion from a thin disc, to a maximum spin parameter of the BH $\hat a=0.9$. 
This value was chosen in agreement with \cite{Gammieetal2004} simulations, which
suggest that the maximum spin MBHs can achieve by coupling with discs in magneto-hydrodynamical 
simulations is $\hat a\simeq 0.9$. 

In the second case, we focus only on a subset of halos, those with effective atomic cooling. 
The cooling curve of metal--free gas has a sharp break at $T < 10^4$K, so that for halos with 
$T_{\rm vir} < 10^4$K the only available coolant is molecular hydrogen. ${\rm H_2}$ is nevertheless 
a rather inefficient coolant \citep{Madauferrararees}. Metal--free halos with virial temperatures 
$T_{\rm vir} > 10^4$K can instead cool even in the absence of ${\rm H_2}$ via neutral hydrogen atomic 
lines to $\sim 8000$ K.  Following \cite{OhHaiman2002}, we assume that a fraction $f_d$ of
the gas settles into an isothermal, exponential disc, embedded in a dark matter halo described by a
\cite{NFW1997} density profile.  The mass of the disc can therefore be expressed as
$M_{\rm disc}= f_d (\Omega_b/\Omega_M) M_h$.  The disc is geometrically thick and it has a very high central density \citep[cfr][] {brommloeb} . We refer the reader to  \cite{VolonteriRees2005} and \cite{LodatoNatarajan} for a detailed description of the model. 

We estimate the mass accreted by the MBH from the surrounding disc within the Bondi-Hoyle formalism \citep{BondiHoyle1944}. 
\begin{eqnarray}
\nonumber
\dot{M}_{\rm Bondi}&=&\frac{\alpha \,4\pi\, G^2\, 
M_{\rm BH}^2\,\mu m_{\rm H} n}{c_s^3}=4\times 10^{-5}\left(\frac{M_{\rm BH}}{10^3\msun}\right)^2\\
&&
\hspace{-0.3cm}
\left(\frac{n_0}{10^4{cm^{-3}}}\right)
\left(\frac{T_\gas}{8000{\rm K}}\right)^{-1.5}\msun\,{\rm yr}^{-1},
\label{eq:dotm}
\end{eqnarray}
where $n_0$ is the density of the gas, of order $10^3{\rm cm^{-3}}\lesssim n_0\lesssim 10^5{\rm cm^{-3}}$
at the center of the gas discs in high redshift halos. 

The collapsing gas disc likely rotates as a rigid body, rotation is therefore small during the initial collapse phase, and the infall of gas is quasi radial. The size of the accretion 
disc, $r_{\rm in}$, is of order of the trapping radius: 
\beq
r_{\rm tr}=r_S\frac{\dot M}{\dot{M}_{\rm Edd}}\propto r_{\rm in}\, M_{\rm BH}^{-1}, 
\label{r_tr}
\eeq
i.e. the radius at which radiation is trapped as the infall speed of the gas is larger than the 
diffusion speed of the radiation. \cite{Begelman1979} and \cite{BegelmanMeier1982} studied 
super-critical accretion onto a BH in spherical geometry and quiescent thick discs 
respectively. Despite the uncertainties, it still seems  possible that when the inflow rate is super-critical, 
the radiative efficiency drops so that the hole can accept the material without greatly exceeding the Eddington luminosity.  The  efficiency could be low either because most radiation is trapped and advected inward, 
or because the flow adjusts so that the material can plunge in from an orbit with small 
binding energy \citep{AbramowiczLasota1980}.

The accretion rate is initially super-critical by a factor of 10 and grows up to a factor of 
about $10^4$ (Volonteri \& Rees 2005), thus making the flow more and more spherical (see 
Equation \ref{r_tr}). On the other hand, the radius of the accretion disc increases steeply with the 
hole mass, until the whole 'plump' accretion disc grows in size enough to cross the trapping 
surface. We assume here that, when the radius of the accretion disc becomes a factor of 5 
larger (although the choice of the exact value is somehow arbitrary) an outflow develops, 
blowing away the disc. Typically, assuming this threshold,  a MBH  accretes only a fraction (1-10\%) of the mass in the disc. Also in this case we follow the evolution of MBH spins as described in \cite{Volonterietal2005}. 

An alternative model for super-critical accretion on seed MBHs can be 
found in \cite{BegelmanVolonteriRees2006}, where unstable discs are considered, instead of
stable systems. Both models for supercritical accretion apply only to metal-free halos with 
virial temperature  $T_{\rm vir} > 10^4$K, that is,  to rare massive halos, at the cosmic time considered. 
Rapid early growth, therefore, can happen only for a tiny fraction of MBH seeds, in a selective and 
biased way.
For illustrative purposes, Figure \ref{fig1} shows how different accretion rates, or the
same accretion rate with different radiative efficiencies can dramatically modify the mass 
of a MBH starting accreting at $z=24$.
\begin{figurehere}
\vspace{0.5cm}
\centerline{\psfig{file=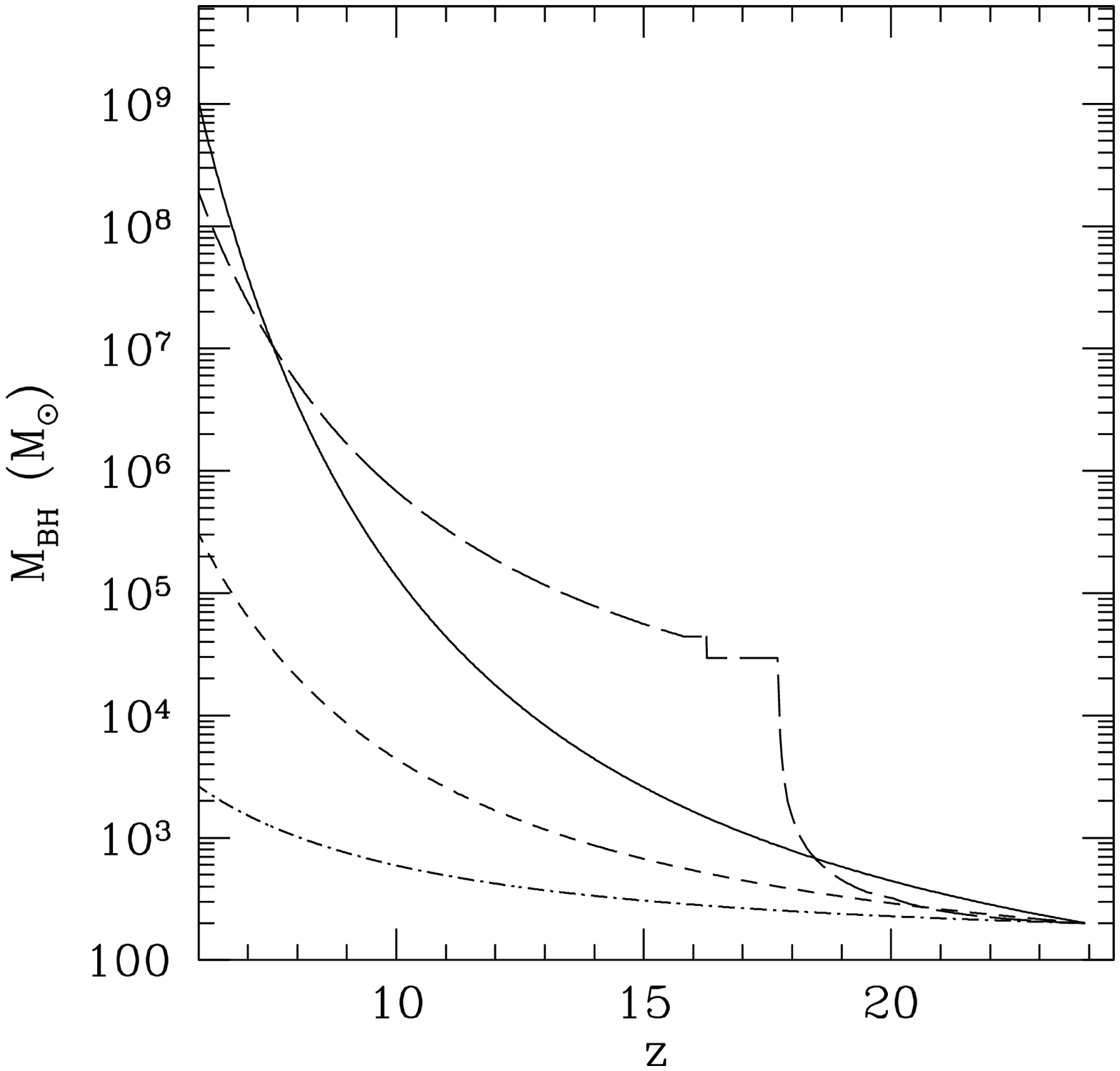,width=3.0in}}
\caption{ Growth of a MBH mass under different assumption for the accretion
rate and efficiency. Eddington limited accretion: $\epsilon=0.1$ ({\it solid line}),
$\epsilon=0.2$ ({\it short dashed line}), $\epsilon=0.4$ ({\it dot-dashed line}). 
Super-critical accretion, as in Volonteri \& Rees 2005 ({\it long dashed line}).}
\label{fig1}
\vspace{0.5cm}
\end{figurehere}

\section{Dynamical evolution  of MBH\lowercase{s}}
To include in this study the dynamics of MBHs, we will follow the evolution 
of MBHs and their hosts in full detail with a semi-analytical technique 
\citep{VHM,Volonterietal2005}.
\subsection{MBH binaries merging efficiency}
The merging -- driven by dynamical friction against the dark matter -- of two
comparable-mass halo$+$MBH systems (``major mergers'') drags in the
satellite hole towards the center of the more massive progenitor,
leading to the formation of a bound MBH binary with separation of
$\sim$ pc.  In massive galaxies at low redshift, the subsequent evolution of the 
binary may be largely determined by three-body interactions with background stars
\citep{BBR1980}. Dark matter particles will be ejected by 
decaying binaries in the same way as the stars. 
Another possibility is that gas processes, rather than three-body interactions with stars or DM, may induce MBH binaries to shrink rapidly and coalesce (e.g. \cite{Mayeretal2006, Dottietal2006, Escalaetal2004, Liu2003, ArmitageNarajan2005, GouldRix2000}). If stellar dynamical or gaseous processes drive the binary sufficiently close ($\lta 0.01$ pc), gravitational radiation will eventually dominate angular momentum and energy losses and cause the
two MBHs to coalesce. 

In gas rich high redshift halos, the orbital evolution of the central SMBH is likely dominated by dynamical friction against the surrounding gaseous medium. The available simulations \citep{Mayeretal2006, Dottietal2006, Escalaetal2004} show that the binary can shrink to about parsec or slightly subparsec scale by dynamical friction against the gas, depending on the gas thermodynamics. These binary separations are still too large for the binary to coalesce within the Hubble time owing to emission of gravitational waves. On the other hand, the interaction between a binary and an accretion disc can lead to a very efficient transport of angular momentum, and the secondary MBH can reach the very subparsec separations at which emission of gravitational radiation dominates on short timescales \cite{Liu2003, ArmitageNarajan2005, GouldRix2000}.  The viscous timescale depends on the properties of the accretion disc and of the binary:
\beq
t_{\rm vis}=0.1\,{\rm Gyr}\,a_1^{3/2}\left(\frac{h}{r}\right)^{-2}_{0.1}\alpha^{-1}_{0.1}\left(\frac{m_1}{10^4 M_\odot}\right)^{-1/2},
\eeq
where $a_1$ is the initial separation of the binary when the secondary MBH starts interacting with the accretion disc in units of parsec, $(h/r)$ is the aspect ratio of the accretion disc, $h/r=0.1$ above, $\alpha$ is the Shakura \& Sunyaev viscosity parameter, $\alpha=0.1$ above, and $m_1$ is the mass of the primary MBH, in solar masses.
The emission of gravitational waves takes over the viscous timescales at a separation \citep{ArmitageNarajan2005}:
\beq
a_{GW}=10^{-8}\, \rm {pc} \left(\frac{h}{r}\right)^{-16/5}_{0.1}\alpha_{0.1}^{-8/5}q_{0.1}^{3/5}\left(\frac{m_1}{10^4 M_\odot}\right),
\eeq
where  $q=m_2/m_1\lta1$,  is the binary mass-ratio. The timescale for coalescence by emission of gravitational waves from $a_{GW}$ is much shorter than the Hubble time:
\beq
t_{\rm gr}={5c^5 a^4(t)\over 256 G^3 m_1m_2(m_1+m_2)}.
\eeq

We have assumed here that, if an accretion disc is surrounding a hard MBH binary, it coalesces  instantaneously owing to interaction with the gas disc. If instead there is no gas readily available, the binary will be losing orbital energy to the stars, if the initial mass function of PopIII stars is bimodal and a large population 
of low mass stars is formed \citep{NakamuraUmemura2001} or to dark matter background otherwise.
The coalescence timescale typically becomes much longer than the Hubble time 
(see \cite{Madauetal2004} for a thorough discussion for this case).  

To test the influence of the merging efficiency, we have compared the above \emph{merging efficient} model to a conservative \emph{merging inefficient} case in which the interaction with gas is neglected, and the binaries shrink only via DM+stars scattering, under the assumption that the loss cone stays full. 

\subsection{Gravitational recoil}
 At high redshift, the recoil as a result of the  
non-zero net linear momentum carried away by gravitational waves, may cause the ejection of MBHs
from the shallow potential wells of their hosts  \citep[e.g.,][]{Madauetal2004,MadauQuataert2004, Merrittetal2004}.   
The recoil velocity has still large uncertainties. Early calculations in the Newtonian regime 
\citep{Fitchett1983}, predict at most a recoil velocity  of $\approx 100 \kms$. 
The Newtonian calculations, in the circular case, predict the center of mass to move on 
circular orbits, spiraling outward while the binary orbit spirals inward, with a velocity
\beq
v_{CM}=1480 \kms \left[\frac{f(q)}{f_{\rm max}}\right]\left(\frac {R_S}{R_L}\right)^{4},
\eeq
where $R_S=2G(m_1+m_2)/c^2$ is the Schwarzschild radius of the system, and the scaling function reads $f(q)/f_{\rm max}$, 
with $f(q)=q^2(1-q)/(1+q)^5$, and $f_{\rm max}=0.01789$.

Here $R_L$ represents the radius 
of the last stable circular orbit, which is $R_L=6G m_1/c^2=3R_Sm_1/(m_1+m_2)$
in the Schwarzschild metric.

The recoil velocity during the plunge phase probably has the largest contribute. 
Calculations of the gravitational recoil inside the innermost
stable orbit (ISCO) naturally have large uncertaintes, but the recoil velocity should be 
constrained between the upper- and lower-limits ($V_{\rm upper}$ and $V_{\rm lower}$ respectively)
suggested by Favata et al. (2004), which span a range between a few $\rm{km\,s^{-1}}$ and 
several hundreds $\rm{km\,s^{-1}}$ for a binary with a mass ratio $q\simeq 0.1$.

\cite{Blanchetetal2005} calculated the gravitational recoil at the second post-Newtonian 
order for non-spinning holes.  Their calculations, available at the moment for Schwarzschild holes only, 
narrow the uncertainties on the final velocity from more than an order of
magnitude \citep{Favataetal2004}, down to about $50\%$. 
\cite{Gopu2006}, using the effective one-body approach, predict velocities about a factor 
of 3 less than  Blanchet et al. The uncertainties quoted in  \cite{Blanchetetal2005} and 
\cite{Gopu2006} make the results of the two calculations incompatible. 

The latest estimate on the recoil comes from fully relativistic numerical simulations \citep{Bakeretal2006} following
the dynamical evolution of a black hole binary within the ISCO. These simulations, carried for a mass ratio $q=0.2$, 
predict a recoil midway through the \cite{Blanchetetal2005} and \cite{Gopu2006} predictions. The recoil predicted by \cite{Bakeretal2006} is still large enough to eject the merging binary from small pre-galactic structures.  As shown by Yoo \& Miralda-Escud\'e 2004,  if the recoil effect is mild (e.g. \cite{Gopu2006,Favataetal2004} lower limits), it is easy to fulfill the constraints on  quasars at $z=6$\footnote{ We performed one calculation adopting the \cite{Favataetal2004} lower limits and checked that
the fraction of displaced binaries amounts to about $5\%$.}. 
As we are considering here the most pessimistic (although realistic) conditions under which it is still 
possible to develop the population of $z=6$ quasars, we adopt as a zeropoint the  \cite{Bakeretal2006}
results. We rescale the recoil to different mass ratios adopting Fitchett's scaling function. Although Fitchett derived
the scaling in the perturbative regime, outside the strong gravity region, both  \cite{Blanchetetal2005} and 
\cite{Gopu2006} find a dependence on the mass ratio in very good agreement with Fitchett's scaling. 

\begin{figurehere}
\vspace{0.5cm}
\centerline{\psfig{file=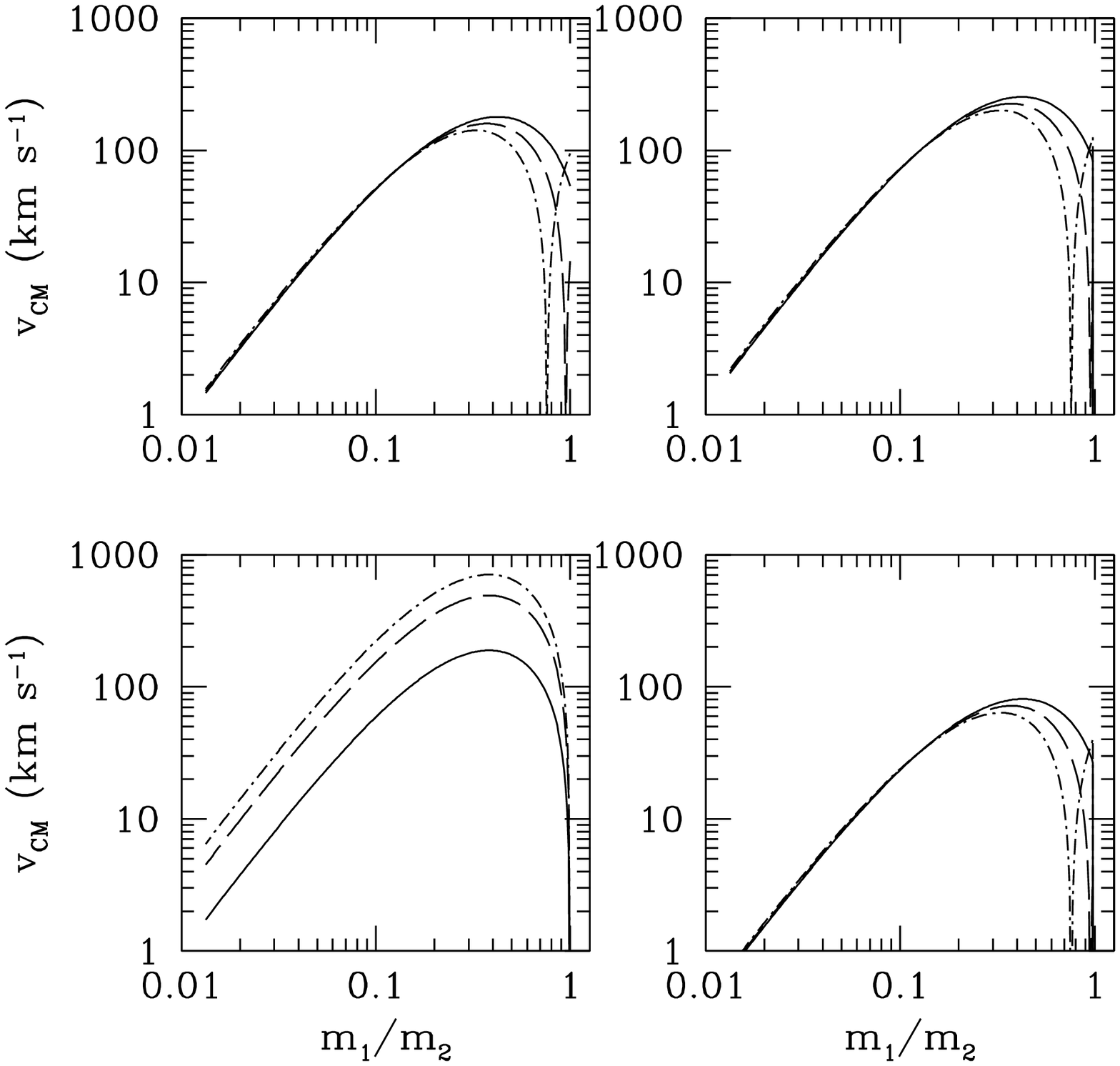,width=3.0in}}
\caption{ Merging BHs recoil velocity as a function of binary mass-ratio. {\it Solid line}: spin parameter $\hat a=1$.
{\it Long-dashed line}: spin parameter $\hat a=0$ .  {\it Dot-dashed line}: spin parameter $\hat a= -1$. 
{\it Bottom-left}: \cite{Favataetal2004} upper limits. 
{\it Top-left}: \cite{Bakeretal2006}. The mass ratio dependence is estimated by the Fitchett's formula, and the 
spin-orbit interaction as in  \cite{Favataetal2004}. 
{\it Top-right}: \cite{Blanchetetal2005}. 
{\it Bottom-right}: \cite{Gopu2006}}
\label{fig2}
\vspace{0.5cm}
\end{figurehere}

One remaining issue is the effect of black hole spins, and the spin-orbit coupling, which however, seems to be mild.  \cite{Favataetal2004} suggest a modification of Fitchett's scaling function to  
keep into account spin-orbit coupling: $\tilde f(q)=f(q)|1 + (7/29)\hat{a}_1/(1-q)|/|1+ (7/29)\hat{a}_1/(1-0.127)|$.
Here $\hat a_1$ is the physical spin parameter of the large hole ($S_1=\hat a_1 GM_{\rm BH}/c$, $\le \hat a_1\le 1$), 
and $\hat a$ is the ``effective'' spin parameter of the binary system. Damour suggests ${\hat a} =
{\hat a}_1 (1 + 3q/4)/(1 + q)^2$ in the post-Newtonian limit.  Figure \ref{fig2} compares different theoretical estimates for the recoil velocity. 

We therefore adopt the following expression to parameterize the recoil velocity:
\begin{equation}
\label{eq:upper}
V_{\rm CM}=463\ {\rm km\ s}^{-1} {\tilde f(q)\over f_{\rm max}.}
\end{equation}

If the recoil velocity of the coalescing binary exceeds the escape 
speed $v_{\rm esc}$, the holes will leave the galaxy altogether. 
If instead $v_{\rm CM}<v_{\rm esc}$, the binary recoils within the galaxy and its orbit
will (slowly) decay owing to dynamical friction \citep{MadauQuataert2004, VolonteriPerna2005}.
The escape velocity calculation assumes an NFW halo for the DM component. The baryonic
component is modeled as an isothermal sphere truncated at the radius of the MBH sphere of 
influence, except in the cases where gas can condense in a disc (see section 3). In this 
case the escape velocity is computed assuming an exponential profile for the gaseous disc. 
The escape velocity is between $2.5-5$ times the virial velocity of the host halo. 
Note, however, that if the recoil velocity is larger than about twice the velocity 
dispersion of the host halo,  the dynamical friction timescale for the ejected MBH to return
to the center of the halo is larger than the Hubble time (Madau \& Quataert 2004).

\section{Results}
The combination of a halo merger tree and our semi-analytical scheme
to trÄt the growth of MBHs and their dynamics is a powerful tool for
tracking the evolution of MBHs at early times. 
Our semi-analytical approach is highly idealized, but allows us 
to explore a large range of different scenarios and their consequences on the
evolution of the MBH population at high redshift.
%The relationship between SMBHs and their host halos is still largely unknown
%at such high redshifts, however. Walter et al. (2005) find that the rotational velocity of 
%the gas in the inner 2.5$kpc$ of SDSS 1148+3251 host is $280 \kms$. We therefore
%add to our ensemble of merger trees the history of a halo  with virial 
%mass $M_h>10^{12}\msun$ at $z=6$. This mass corresponds to a virial circular velocity of 
%$250\kms$. 

\subsection{MBH dynamics}
The dynamical and accretion evolution of high-redshift MBHs is so strongly intermingled that it is very difficult to clearly separate the relative effects. The accretion history determines the MBH binaries mass-ratios, and therefore the merging timescale, and the recoil velocity (cfr. eq \ref{eq:upper}). 

\begin{figurehere}
\vspace{0.5cm}
\centerline{\psfig{file=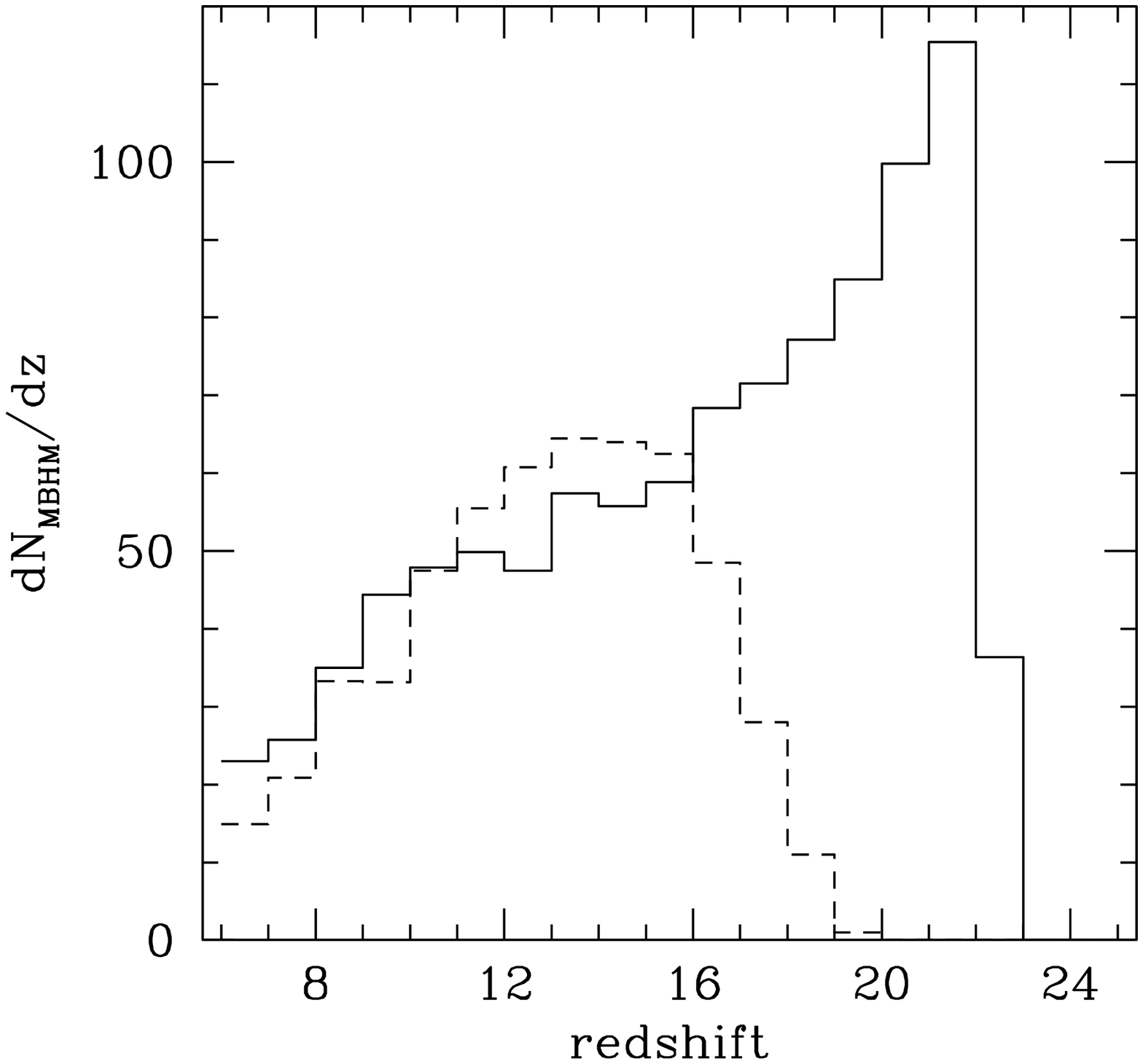,width=3.0in}}
\caption{ Averaged distribution of MBH mergers per merger tree of a $10^{13}\msun$ halo 
at $z=6$. {\it Solid line:} when an accretion disc is present, the satellite MBH is dragged towards the central MBH on the viscous timescale (\emph{efficient merging}). {\it Dashed line:} binaries evolve via DM+stars scattering (\emph{inefficient merging}).}
\vspace{0.5cm}
\label{fig3}
\end{figurehere}

Figure \ref{fig3} compares the MBH merger rates in two different models accounting for the orbital evolution of MBH binaries in the phase preceding emission of gravitational waves. If gas processes (\emph{efficient merging}) 
drive the MBH orbital decay, MBHs start merging efficiently at very early times, when host DM halos 
are still small. Although the absolute number of mergers is larger, more MBHs are ejected from DM halos due to the rocket effect, because halos are smaller at earlier times. The two effects, increased mergers and increased ejections, compensate (see also figure \ref{fig5}) thus leading to similar SMBH masses in the main halo at $z=6$. Note that the 
MBH merger rate in such a highly biased volume is not the average cosmic rate, as we are following only the hierarchy leading to extremely high mass halos at $z=6$ (cfr. \cite{Diemandetal2005} for similar considerations at lower redshift).  MBHs in a high-overdensity volume experience a much higher number of mergers with respect to the typical MBH evolving in an less overdense region. 

\begin{figurehere}
\vspace{0.5cm}
\centerline{\psfig{file=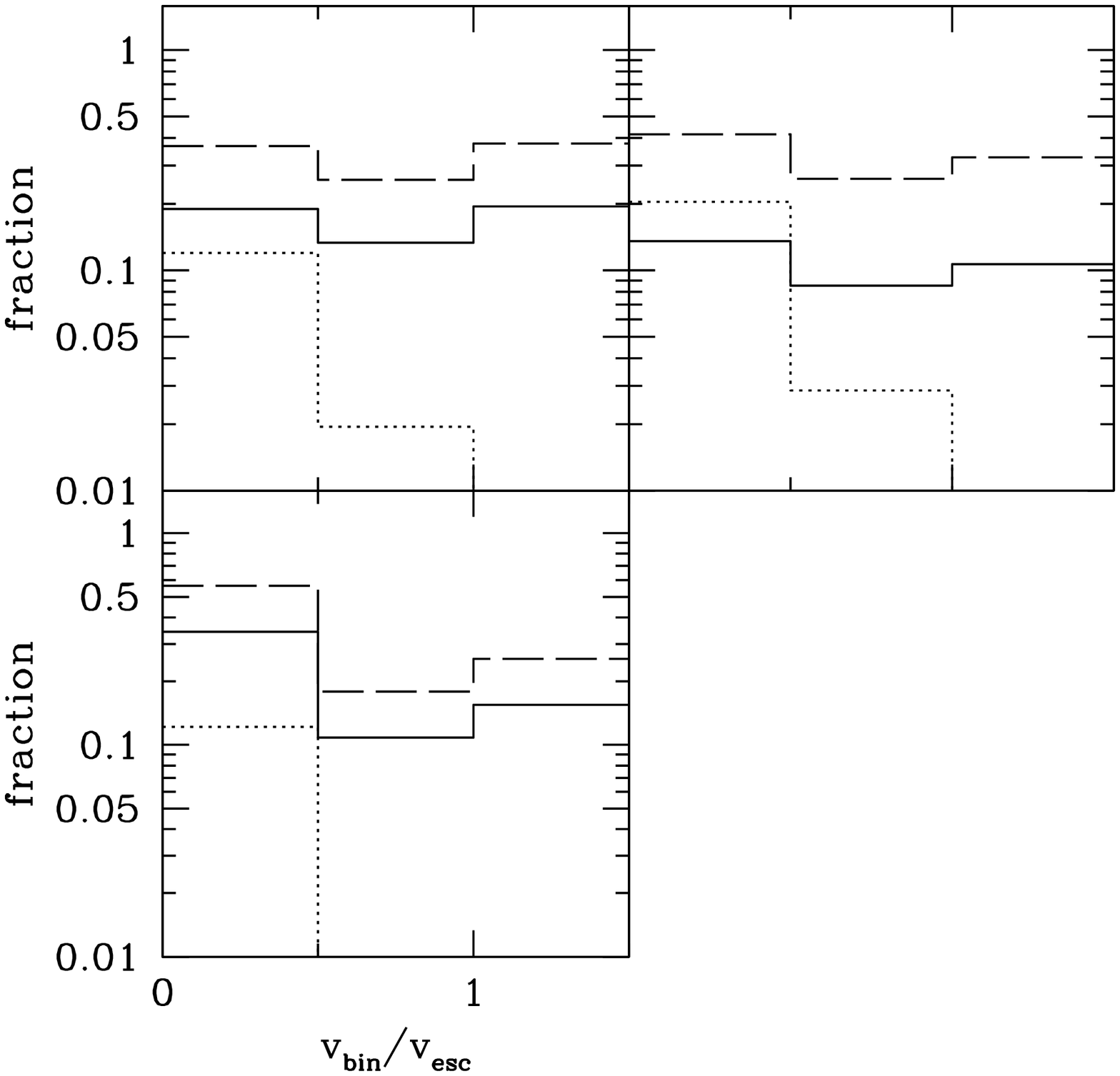,width=3.0in}}
\caption{ Distribution of kicked BHs as a function the 
MBH binary recoil velocity to the escape velocity from the host halo ratio. {\it Solid line:} 
fraction of MBHs with respect to the total number of seeds. {\it Dotted line:}  only for binaries 
merging at $z<10$. {\it Dashed line:} number of MBHs over the number of binaries. The three panels refer to different models. {\it Upper left panel:} Eddington accretion rate only ($\epsilon \leq 0.16$), 
\emph{efficient merging}. {\it Upper right panel:}  Eddington accretion 
rate only ($\epsilon \leq 0.16$), \emph{inefficient merging}. 
{\it Lower left panel}: MBHs are allowed to grow at supercritical rates, thus shifting the mass 
ratios of merging binaries towards mass ratios $\ll 1$. \emph{Efficient merging} is assumed.
%{\it Lower right panel:} Eddington accretion 
%rate only ($\epsilon \leq 0.16$), binaries evolve via DM+stars scattering, initial MBH mass function 
%flat within the mass ranges $20<m_\bullet<70\,\msun$ and $130<m_\bullet<600\,\msun$.
}
\label{fig4}
\vspace{0.5cm}
\end{figurehere}

Figure \ref{fig4} summarizes our results in terms of MBH binaries ejection. We have explored a large 
parameter space and we selected here a few representative examples. Accretion and dynamical
histories are strongly intermingled, as the kick velocity depends on the binary mass-ratio. 
The rocket recoil is substantial only for mass-ratios of order $q\gta 0.1$, which is believed to be 
the most probable mass ratio for MBH binaries (due to the inefficiency of dynamical friction in 
bringing the satellite MBH within the influence of the main one in case of minor mergers, 
see also \cite{Sesanaetal2005}). Interestingly, the Volonteri \& Rees model for supercritical 
accretion skews the mass ratios of MBH binaries towards lower values, as the conditions 
for supercritical accretion to happen are typically fulfilled by only one of two merging 
systems. This latter result supports a biased and selective growth of high-z BHs, as the conditions for super-critical accretion appear to be fulfilled only in halos with $T_{\rm vir} > 10^4$K, which represent 3-$\sigma$ or 4-$\sigma$ peaks in the field of density fluctuations . If only a tiny fraction of the MBHs, those hosted in the most massive halos at this early time, undergo a rapid growth, the conditions for a mild effect of the recoil are naturally met, regardless of which detailed kick velocity calculation is considered. The distribution of merging MBHs mass ratios is in fact skewed towards low mass ratios ( $q=m_2/m_1\ll1$), where the expected kick velocity is low (Figure \ref{fig2}).

\subsection{MBH accretion}
The mass growth history of a MBH ending up as a SMBH in a  $M_h=10^{13}\msun$  at $z=6$ 
is shown in figure \ref{fig5} for a series of different models. MBH mass increase by mergers 
is  taken into account, as well as mergers 'negative feedback' onto the MBH growth due to the possibility of
binary ejection following coalescence. 

\begin{figurehere}
\vspace{0.5cm}
\centerline{\psfig{file=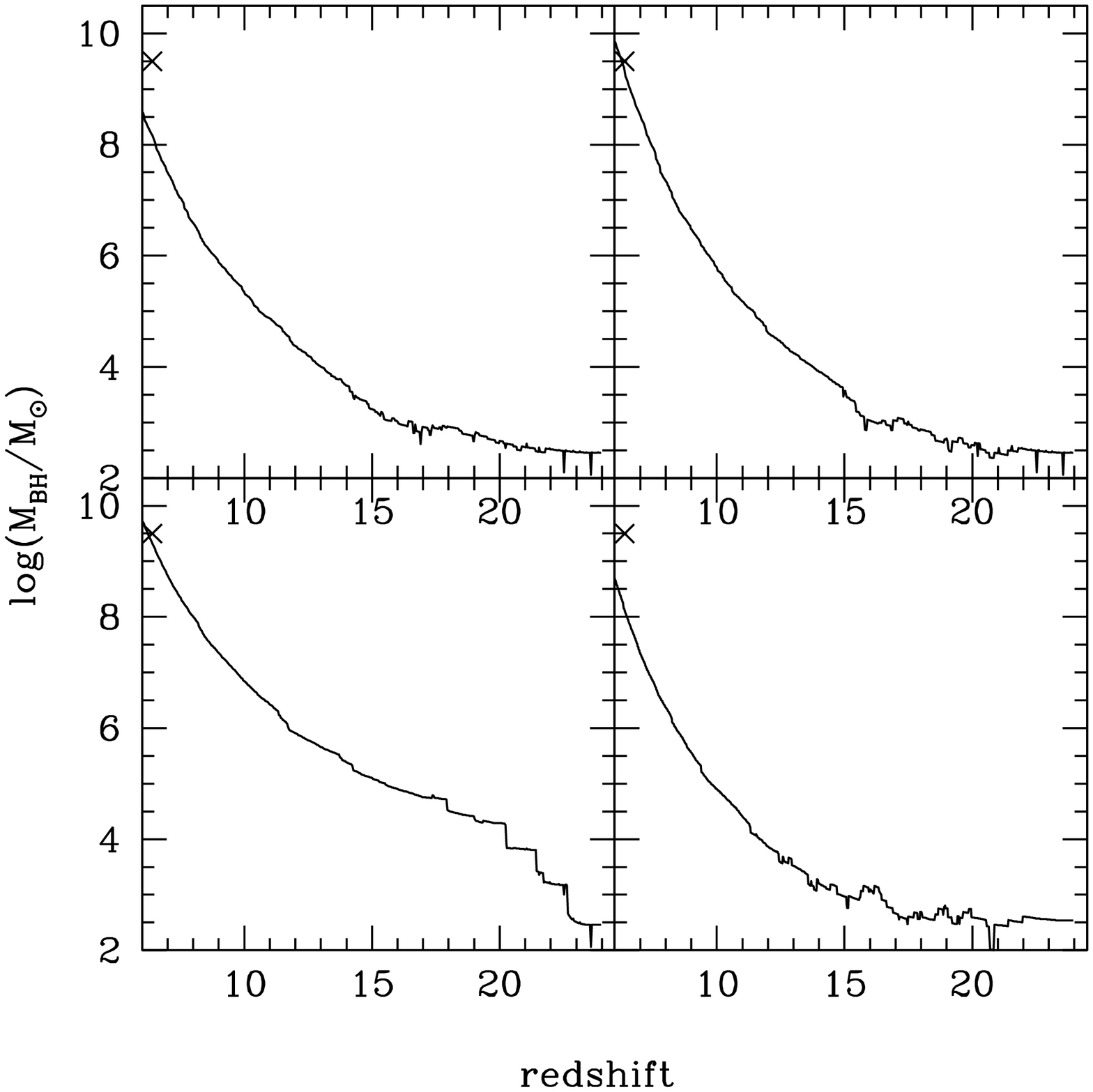,width=3.0in}}
\caption{ Averaged mass growth history of the MBH at the center of a $10^{13}\msun$ halo 
at $z=6$. 
{\it Upper left panel:} Eddington accretion rate only (maximum radiative efficiency $\epsilon=0.16$). 
{\it Upper right panel:} Eddington accretion rate only (maximum radiative efficiency $\epsilon=0.12$). 
{\it Lower left panel:} super-critical accretion, in a fat disc with  $f_d=0.1$.
{\it Lower right panel:} Eddington accretion rate only (maximum radiative efficiency $\epsilon=0.16$), 
and 25 times larger seed holes abundance. A cross marks the locus of SDSS 1148+3251.}
\label{fig5}
\vspace{0.5cm}
\end{figurehere}

As a reference, we have considered an upper limit to the radiative efficiency of $\epsilon=0.16$, or $\hat a=0.9$ adopting the standard conversion, inspired by \cite{Gammieetal2004} simulations.  Figure \ref{fig3} shows, however, that MBH masses of order $10^{9}\msun$ can be reached by $z=6$ only if dynamical effects are not too destructive. In fact, if we adopt the \cite{Bakeretal2006} scaling (Eq. \ref{eq:upper})
for the gravitational recoil velocity, MBH masses are always less than a billion solar masses. 
The constraints imposed by $z=6$ quasars can be met if either (i) the radiative efficiency is 
lower than $\epsilon_{\rm max}=0.12$ (corresponding to a spin parameter $\hat a=0.8$) or (ii) 
if the number of seed MBHs is much more than 25 times\footnote{Computational costs did not allow to explore initial conditions with a larger number of MBH seeds.} larger (i.e. seeds inhabit density peaks 
$\ll3.5-\sigma$ at z=24. Note that in this case massive black holes would be expected also 
in dwarf galaxies with total masses well below $M_h=10^{11}\msun$ today).  
Alternatively, (iii) if the MBHs go through a phase of supercritical accretion.

If the dynamical influence of the 'gravitational rocket' is better described by \cite[e.g.][]{Favataetal2004, Gopu2006}, though, the observational constrains can be met more easily, 
allowing for higher radiative efficiencies or rarer seeds. It is worth noting that a 
radiative efficiency $\epsilon>0.16$ can be accommodated only if the influence of mergers is large
(see also Shapiro 2005 and Yoo \& Miralda-Escud\'e 2004), e.g. seeds in lower $\sigma$-peaks and 
\emph{efficient merging} of binary MBHs.  

\begin{figurehere}
\vspace{0.5cm}
\centerline{\psfig{file=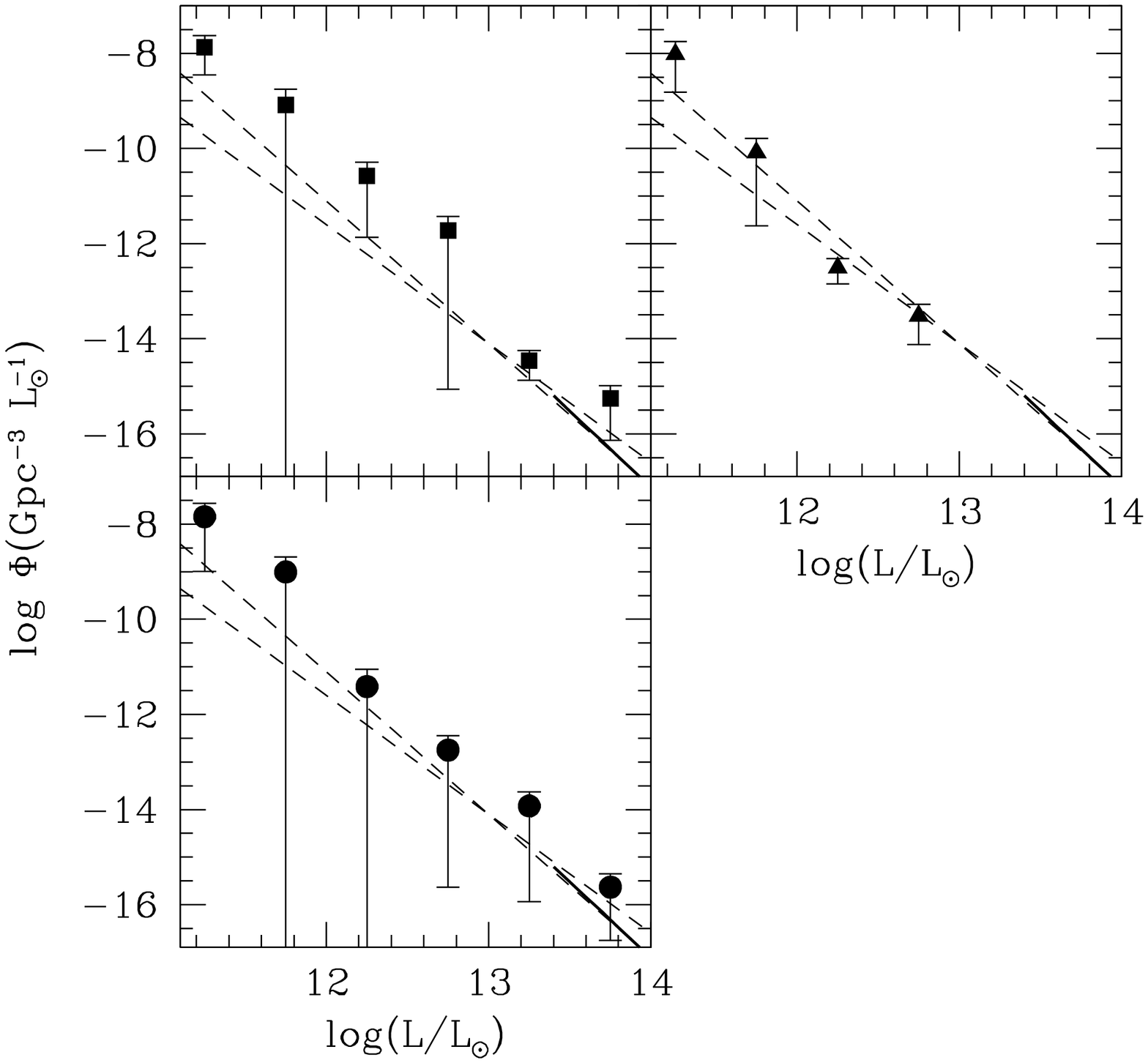,width=3.0in}}
\caption{Luminosity function of $z=6$ quasars, resulting from different models. 
The thick solid line shows the best fit LF from Fan \etal 2004. The dashed lines two model LF 
with faint and bright end slopes of the luminosity function $(\alpha,\beta)=(-1,-2.5)$ 
and $(\alpha,\beta)=(-2,-3.5)$.
{\it Upper left panel:} Eddington accretion rate only (maximum radiative efficiency $\epsilon=0.12$).
{\it Upper right panel:}  Eddington accretion rate only (maximum radiative efficiency $\epsilon=0.16$), 
{\it Lower left panel:} super-critical accretion, in a fat disc with  $f_d=0.1$.  
}
\label{fig6}
\vspace{0.5cm}
\end{figurehere}

We have compared theoretical luminosity functions (LF) with the most recent determination of the 
quasar blue LF from the SDSS \citep{Fanetal2004}. The SDSS samples only the very bright end of the 
LF; \cite{Fanetal2001b} fit a single power-law to the data, which span luminosities larger than 
$10^{13}\,{\rm L_\odot}$ in the blue band. Upper limits on the faint end of the luminosity 
function, though, can be derived from the non-detection of $z=6$ quasars in the 
Chandra Deep Field-North \citep{Bargeretal2003} and in the Canada-France High-z Quasar 
Survey \citep{Willottetal2005}. We have adopted here the parameterization of the LF given
by  \cite{Willottetal2005} and compared our results with two representative cases for the faint 
($\alpha$) and bright ($\beta$) end slopes of the luminosity function: 
$(\alpha,\beta)=(-1,-2.5)$ and $(\alpha,\beta)=(-2,-3.5)$. Letting the luminosity break
vary between $L^*_B=7\times10^{11} L_\odot$ and $L^*_B=2\times10^{12}L\odot$ leads to indistinguishable 
curves. Figure \ref{fig6} shows that quasars bright enough to reproduce the observed LF cannot be
created under the assumption of accretion is limited at the Eddington rate, occurring via a disc
which can spin up black holes up to $\hat a=0.9$ (corresponding using the standard conversion to  $\epsilon=0.16$). 
The constraints from the faint end of LF are still very weak (cfr. Figure 5 in \cite{Willottetal2005}), so
we cannot rule out either a model with a lower $\epsilon_{\rm max}$ or a model with supercritical 
accretion based on these results only. 

\section{Discussion}
The detection of luminous quasars at $z\simeq6$, suggesting an early growth of SMBHs, 
deserves a special investigation within the hierarchical scenario of galaxy formation. 
Following a series of papers tracing the seeds of the SMBH population observed today from 
the first stars, we have here focused on the constraints set by the presence of SMBHs already 
in place when the Universe was less than one billion year old.

The haloes that we choose for our investigation, $M_h>10^{13}\msun$  at $z=6$, are 5.5-$\sigma$ 
density fluctuation. They are not therefore representative of the typical halo mass at $z=6$. 
On the other hand, the masses are chosen requiring that their number density matches the 
number density of quasars derived by Fan et al. (2004). That the highest massive halos 
experienced strong evolution at early times is in line with the 'anti-hierarchical' evolution of 
galactic structures suggested by high-redshift galactic surveys \citep[e.g.][]{Kodamaetal04}.
A simple exercise can help us quantify the global importance of the black holes hosted in the 
selected halos. 

If the correlation between black hole mass and halo circular velocity holds up to $z=6$, we can use 
the Press \& Schechter formalism 
%(taking into account the modification suggested by Jenkins et al. 2001) 
to determine the MBH density as a function of halo mass at $z=6$. The MBH mass 
density in halos with mass larger then $M_h=10^{13}\msun$ is $\simeq 20\msun {\rm \,Mpc^{-3}}$, while 
for $M_h=10^{12}\msun$ we have $\simeq 10^3\msun {\rm \,Mpc^{-3}}$, and for $M_h=10^{11}\msun$ we have 
$\simeq 10^4\msun {\rm \,Mpc^{-3}}$. So, if all halos with, e.g.,  $M_h=10^{11}\msun$, at $z=6$ host a SMBH whose 
mass scales with the local  $M_{\rm BH}-V_c$ , the SMBH density at 
$z=6$ is similar to that at $z=3$, leaving very small room for accretion between $z=6$ and $z=3$. 
This implies that either the $M_{\rm BH}-V_c$ relation is redshift dependent,
and in the opposite sense than suggested by the \cite{Walteretal2004} observation, or the occupation fraction
of black holes is not unity at all redshift. We define the MBH occupation fraction, as the average number
of halos hosting a MBH over the total number of halos in a comoving volume. 
Volonteri, Haardt \& Madau 2003 (see also \cite{Marulli2006}) show that models in which SMBHs evolve from seeds forming in biased regions (e.g., remnants of PopIII stars), predict a SMBH occupation fraction which decreases with halo mass. 
%The resulting occupation fraction, can be fit as ${\rm OF}=-0.125*\log(M_h)^2+3.075\log(M_h)-17.9$ 
%for $3\times10^9\msun<M_h<3\times10^{12}\msun$. 

If the occupation fraction is a declining function of the halo mass, we expect MBH-MBH interactions to be 
more common for massive halos. MBHs hosted in low mass halos grow \emph{in isolation}, while 
MBHs lurking in massive halos have a much higher probability of experiencing a MBH-MBH merger. 
So, on one hand MBHs hosted in low mass halos have a larger probability of being ejected by a kick, on the 
other hand their probability of experiencing a merger in the first place is smaller. 
We might therefore expect that a SMBH can grow almost undisturbed by $z=6$ up to a billion 
solar masses, if hosted in a, say, $10^{11}\msun$ halo. A requirement in this case is, however, that
the mass of the black holes, in these high redshift systems, can grow to masses much larger than 
allowed by the local $M_{\rm BH}-V_c$ or  correlation. This is the opposite of what seen in numerical 
simulations \citep{Robertson2005}, but in agreement with the \cite{Walteretal2004} results. 
%It might indeed be 
%possible that at these very high redshift, before strong stellar bulges are formed, SMBHs masses correlate with 
%the mass of the halo (see also Marulli et al. 2006). 

Summarizing, if the SMBHs powering $z=6$ quasars are hosted in halos with $M_h\lta10^{12}\msun$, they can grow 
almost \emph{in isolation}, and accretion can be limited to the Eddington rate, provided the radiative efficiency is not too 
high, but the mass of the SMBHs must be allowed to exceed the nominal $M_{BH}-$velocity dispersion correlation. 

If the host halo masses are of order $M_h\simeq10^{13}\msun$, dynamical interactions are important in their evolutionary 
history. In this case, either kick velocities are at the lower end of the predicted range, or accretion
occurs at supercritical rates for some periods of time. 

%If accretion is always limited by the Eddington rate via a thin disc (i.e., with a
%high efficiency of 'spin-up', see Volonteri \etal 2005), MBH masses larger than
%$10^9\, \msun$ can be reached only if the gravitational recoil is not important ('lower limits'
%to the recoil velocity estimated by Favata \etal). In the latter  case, the maximum radiative 
%efficiency allowing to reproduce the LF at $z=6$ is $\epsilon=0.12$. 

If, instead, high-redshift MBHs can accrete at super-critical rate during an early phase, reproducing the 
observed MBH mass values is not an issue, even in the case that the recoil velocity is in the 
'upper limits' range. Similar conclusion can be reached also with alternative models for early 
supercritical accretion onto MBH seeds formed out of direct collapse of gas in high-redshift halos 
(Begelman, Volonteri \& Rees, 2006). 

The emerging picture, therefore, points to a Darwinian natural selection scenario. At early times only a small fraction of MBHs, those hosted in the highest density peaks ($T_{\rm vir} > 10^4$K), evolve rapidly and efficiently (see also \cite{Jbromm}). The effects are twofold: not only these rare BHs become quickly supermassive owing to efficient accretion, 
but also the effects of dangerous dynamical interactions are softened, as the merging of low
mass ratio MBH binaries is favored. Alternative models for MBH seed formation \citep[e.g.][]{kous2004} tend to select 
as sites of black hole formation halos with $T_{\rm vir} > 10^4$K as well.

Clarification of the early evolution of SMBHs can come from detection of the gravitational wave signal 
from their inspiral, as different models for the MBH early evolution predict different mass-ratio
distributions for the merging MBHs. Reliable calculations narrowing the uncertainties on 
the kick velocity due to gravitational emission for spinning MBH binaries will help clarify the threat that kicks
represent for the early evolution of the MBH population.
\smallskip

\acknowledgements 
We have benefitted from useful discussions with Piero Madau, Francesco Haardt, Cole Miller, and Monica Colpi. MV thanks the Department of Astronomy \& Astrophysics of the University
of California at Santa Cruz for the warm hospitality. We would like to thank the referee for helpful comments.

\bibliography{paper.bib}

\begin{thebibliography}{61}
\expandafter\ifx\csname natexlab\endcsname\relax\def\natexlab#1{#1}\fi

\bibitem[{{Abramowicz} \& {Lasota}(1980)}]{AbramowiczLasota1980}
{Abramowicz}, M.~A. \& {Lasota}, J.~P. 1980, Acta Astronomica, 30, 35

\bibitem[{{Aller} \& {Richstone}(2002)}]{Aller2002}
{Aller}, M.~C. \& {Richstone}, D. 2002, \aj, 124, 3035

\bibitem[{{Armitage} \& {Natarajan}(2005)}]{ArmitageNarajan2005}
{Armitage}, P.~J. \& {Natarajan}, P. 2005, \apj, 634, 921

\bibitem[{{Baker} {et~al.}(2006){Baker}, {Centrella}, {Choi}, {Koppitz}, {van
  Meter}, \& {Miller}}]{Bakeretal2006}
{Baker}, J.~G., {Centrella}, J., {Choi}, D.-I., {Koppitz}, M., {van Meter},
  J.~R., \& {Miller}, M.~C. 2006, ArXiv Astrophysics e-prints

\bibitem[{{Barger} {et~al.}(2003){Barger}, {Cowie}, {Capak}, {Alexander},
  {Bauer}, {Brandt}, {Garmire}, \& {Hornschemeier}}]{Bargeretal2003}
{Barger}, A.~J., {Cowie}, L.~L., {Capak}, P., {Alexander}, D.~M., {Bauer},
  F.~E., {Brandt}, W.~N., {Garmire}, G.~P., \& {Hornschemeier}, A.~E. 2003,
  \apjl, 584, L61

\bibitem[{{Barth} {et~al.}(2003){Barth}, {Martini}, {Nelson}, \&
  {Ho}}]{Barthetal2003}
{Barth}, A.~J., {Martini}, P., {Nelson}, C.~H., \& {Ho}, L.~C. 2003, \apjl,
  594, L95

\bibitem[{{Begelman}(1979)}]{Begelman1979}
{Begelman}, M.~C. 1979, \mnras, 187, 237

\bibitem[{{Begelman} {et~al.}(1980){Begelman}, {Blandford}, \&
  {Rees}}]{BBR1980}
{Begelman}, M.~C., {Blandford}, R.~D., \& {Rees}, M.~J. 1980, \nat, 287, 307

\bibitem[{{Begelman} \& {Meier}(1982)}]{BegelmanMeier1982}
{Begelman}, M.~C. \& {Meier}, D.~L. 1982, \apj, 253, 873

\bibitem[{{Begelman} {et~al.}(2006){Begelman}, {Volonteri}, \&
  J.}]{BegelmanVolonteriRees2006}
{Begelman}, M.~C., {Volonteri}, M., \& J., R.~M. 2006, MNRAS, submitted,
  astro-ph/0602363

\bibitem[{{Blanchet} {et~al.}(2005){Blanchet}, {Qusailah}, \&
  {Will}}]{Blanchetetal2005}
{Blanchet}, L., {Qusailah}, M.~S.~S., \& {Will}, C.~M. 2005, \apj, 635, 508

\bibitem[{{Bondi} \& {Hoyle}(1944)}]{BondiHoyle1944}
{Bondi}, H. \& {Hoyle}, F. 1944, \mnras, 104, 273

\bibitem[{{Bromm} \& {Loeb}(2003)}]{brommloeb}
{Bromm}, V. \& {Loeb}, A. 2003, \apj, 596, 34

\bibitem[{{Cox}(2004)}]{tjphd}
{Cox}, T.~J. 2004, Ph.D.~Thesis

\bibitem[{{Damour} \& {Gopakumar}(2006)}]{Gopu2006}
{Damour}, T. \& {Gopakumar}, A. 2006, ArXiv General Relativity and Quantum
  Cosmology e-prints

\bibitem[{{Di Matteo} {et~al.}(2005){Di Matteo}, {Springel}, \&
  {Hernquist}}]{DiMatteo2005}
{Di Matteo}, T., {Springel}, V., \& {Hernquist}, L. 2005, \nat, 433, 604

\bibitem[{{Diemand} {et~al.}(2005){Diemand}, {Madau}, \&
  {Moore}}]{Diemandetal2005}
{Diemand}, J., {Madau}, P., \& {Moore}, B. 2005, \mnras, 364, 367

\bibitem[{{Dotti} {et~al.}(2006){Dotti}, {Colpi}, \& {Haardt}}]{Dottietal2006}
{Dotti}, M., {Colpi}, M., \& {Haardt}, F. 2006, \mnras, 367, 103

\bibitem[{{Escala} {et~al.}(2004){Escala}, {Larson}, {Coppi}, \&
  {Mardones}}]{Escalaetal2004}
{Escala}, A., {Larson}, R.~B., {Coppi}, P.~S., \& {Mardones}, D. 2004, \apj,
  607, 765

\bibitem[{{Fan} {et~al.}(2004){Fan}, {Hennawi}, {Richards}, {Strauss},
  {Schneider}, {Donley}, {Young}, {Annis}, {Lin}, {Lampeitl}, {Lupton}, {Gunn},
  {Knapp}, {Brandt}, {Anderson}, {Bahcall}, {Brinkmann}, {Brunner}, {Fukugita},
  {Szalay}, {Szokoly}, \& {York}}]{Fanetal2004}
{Fan}, X., {Hennawi}, J.~F., {Richards}, G.~T., {Strauss}, M.~A., {Schneider},
  D.~P., {Donley}, J.~L., {Young}, J.~E., {Annis}, J., {Lin}, H., {Lampeitl},
  H., {Lupton}, R.~H., {Gunn}, J.~E., {Knapp}, G.~R., {Brandt}, W.~N.,
  {Anderson}, S., {Bahcall}, N.~A., {Brinkmann}, J., {Brunner}, R.~J.,
  {Fukugita}, M., {Szalay}, A.~S., {Szokoly}, G.~P., \& {York}, D.~G. 2004,
  \aj, 128, 515

\bibitem[{{Fan} {et~al.}(2001{\natexlab{a}}){Fan}, {Narayanan}, {Lupton},
  {Strauss}, {Knapp}, {Becker}, {White}, {Pentericci}, {Leggett}, {Haiman},
  {Gunn}, {Ivezi{\'c}}, {Schneider}, {Anderson}, {Brinkmann}, {Bahcall},
  {Connolly}, {Csabai}, {Doi}, {Fukugita}, {Geballe}, {Grebel}, {Harbeck},
  {Hennessy}, {Lamb}, {Miknaitis}, {Munn}, {Nichol}, {Okamura}, {Pier},
  {Prada}, {Richards}, {Szalay}, \& {York}}]{Fanetal2001b}
{Fan}, X., {Narayanan}, V.~K., {Lupton}, R.~H., {Strauss}, M.~A., {Knapp},
  G.~R., {Becker}, R.~H., {White}, R.~L., {Pentericci}, L., {Leggett}, S.~K.,
  {Haiman}, Z., {Gunn}, J.~E., {Ivezi{\'c}}, {\v Z}., {Schneider}, D.~P.,
  {Anderson}, S.~F., {Brinkmann}, J., {Bahcall}, N.~A., {Connolly}, A.~J.,
  {Csabai}, I., {Doi}, M., {Fukugita}, M., {Geballe}, T., {Grebel}, E.~K.,
  {Harbeck}, D., {Hennessy}, G., {Lamb}, D.~Q., {Miknaitis}, G., {Munn}, J.~A.,
  {Nichol}, R., {Okamura}, S., {Pier}, J.~R., {Prada}, F., {Richards}, G.~T.,
  {Szalay}, A., \& {York}, D.~G. 2001{\natexlab{a}}, \aj, 122, 2833

\bibitem[{{Fan} {et~al.}(2001{\natexlab{b}}){Fan}, {Strauss}, {Schneider},
  {Gunn}, {Lupton}, {Becker}, {Davis}, {Newman}, {Richards}, {White},
  {Anderson}, {Annis}, {Bahcall}, {Brunner}, {Csabai}, {Hennessy}, {Hindsley},
  {Fukugita}, {Kunszt}, {Ivezi{\'c}}, {Knapp}, {McKay}, {Munn}, {Pier},
  {Szalay}, \& {York}}]{Fanetal2001a}
{Fan}, X., {Strauss}, M.~A., {Schneider}, D.~P., {Gunn}, J.~E., {Lupton},
  R.~H., {Becker}, R.~H., {Davis}, M., {Newman}, J.~A., {Richards}, G.~T.,
  {White}, R.~L., {Anderson}, J.~E., {Annis}, J., {Bahcall}, N.~A., {Brunner},
  R.~J., {Csabai}, I., {Hennessy}, G.~S., {Hindsley}, R.~B., {Fukugita}, M.,
  {Kunszt}, P.~Z., {Ivezi{\'c}}, {\v Z}., {Knapp}, G.~R., {McKay}, T.~A.,
  {Munn}, J.~A., {Pier}, J.~R., {Szalay}, A.~S., \& {York}, D.~G.
  2001{\natexlab{b}}, \aj, 121, 54

\bibitem[{{Favata} {et~al.}(2004){Favata}, {Hughes}, \&
  {Holz}}]{Favataetal2004}
{Favata}, M., {Hughes}, S.~A., \& {Holz}, D.~E. 2004, \apjl, 607, L5

\bibitem[{{Ferrarese}(2002)}]{Ferrarese2002}
{Ferrarese}, L. 2002, \apj, 578, 90

\bibitem[{{Fitchett}(1983)}]{Fitchett1983}
{Fitchett}, M.~J. 1983, \mnras, 203, 1049

\bibitem[{{Gammie} {et~al.}(2004){Gammie}, {Shapiro}, \&
  {McKinney}}]{Gammieetal2004}
{Gammie}, C.~F., {Shapiro}, S.~L., \& {McKinney}, J.~C. 2004, \apj, 602, 312

\bibitem[{{Gould} \& {Rix}(2000)}]{GouldRix2000}
{Gould}, A. \& {Rix}, H.-W. 2000, \apjl, 532, L29

\bibitem[{{Johnson} \& {Bromm}(2006)}]{Jbromm}
{Johnson}, J.~L. \& {Bromm}, V. 2006, ArXiv Astrophysics e-prints

\bibitem[{{Kazantzidis} {et~al.}(2005){Kazantzidis}, {Mayer}, {Colpi}, {Madau},
  {Debattista}, {Wadsley}, {Stadel}, {Quinn}, \& {Moore}}]{Steliosetal2005}
{Kazantzidis}, S., {Mayer}, L., {Colpi}, M., {Madau}, P., {Debattista}, V.~P.,
  {Wadsley}, J., {Stadel}, J., {Quinn}, T., \& {Moore}, B. 2005, \apjl, 623,
  L67

\bibitem[{{Kodama} {et~al.}(2004){Kodama}, {Yamada}, {Akiyama}, {Aoki}, {Doi},
  {Furusawa}, {Fuse}, {Imanishi}, {Ishida}, {Iye}, {Kajisawa}, {Karoji},
  {Kobayashi}, {Komiyama}, {Kosugi}, {Maeda}, {Miyazaki}, {Mizumoto},
  {Morokuma}, {Nakata}, {Noumaru}, {Ogasawara}, {Ouchi}, {Sasaki}, {Sekiguchi},
  {Shimasaku}, {Simpson}, {Takata}, {Tanaka}, {Ueda}, {Yasuda}, \&
  {Yoshida}}]{Kodamaetal04}
{Kodama}, T., {Yamada}, T., {Akiyama}, M., {Aoki}, K., {Doi}, M., {Furusawa},
  H., {Fuse}, T., {Imanishi}, M., {Ishida}, C., {Iye}, M., {Kajisawa}, M.,
  {Karoji}, H., {Kobayashi}, N., {Komiyama}, Y., {Kosugi}, G., {Maeda}, Y.,
  {Miyazaki}, S., {Mizumoto}, Y., {Morokuma}, T., {Nakata}, F., {Noumaru}, J.,
  {Ogasawara}, R., {Ouchi}, M., {Sasaki}, T., {Sekiguchi}, K., {Shimasaku}, K.,
  {Simpson}, C., {Takata}, T., {Tanaka}, I., {Ueda}, Y., {Yasuda}, N., \&
  {Yoshida}, M. 2004, \mnras, 350, 1005

\bibitem[{{Koushiappas} {et~al.}(2004){Koushiappas}, {Bullock}, \&
  {Dekel}}]{kous2004}
{Koushiappas}, S.~M., {Bullock}, J.~S., \& {Dekel}, A. 2004, \mnras, 354, 292

\bibitem[{{Krolik} {et~al.}(2005){Krolik}, {Hawley}, \&
  {Hirose}}]{Kroliketal2005}
{Krolik}, J.~H., {Hawley}, J.~F., \& {Hirose}, S. 2005, \apj, 622, 1008

\bibitem[{{Liu} {et~al.}(2003){Liu}, {Wu}, \& {Cao}}]{Liu2003}
{Liu}, F.~K., {Wu}, X.-B., \& {Cao}, S.~L. 2003, \mnras, 340, 411

\bibitem[{{Lodato} \& {Natarajan}(2006)}]{LodatoNatarajan}
{Lodato}, G. \& {Natarajan}, P. 2006, ArXiv Astrophysics e-prints

\bibitem[{{Madau} {et~al.}(2001){Madau}, {Ferrara}, \&
  {Rees}}]{Madauferrararees}
{Madau}, P., {Ferrara}, A., \& {Rees}, M.~J. 2001, \apj, 555, 92

\bibitem[{{Madau} \& {Quataert}(2004)}]{MadauQuataert2004}
{Madau}, P. \& {Quataert}, E. 2004, \apjl, 606, L17

\bibitem[{{Madau} \& {Rees}(2001)}]{MadauRees2001}
{Madau}, P. \& {Rees}, M.~J. 2001, \apjl, 551, L27

\bibitem[{{Madau} {et~al.}(2004){Madau}, {Rees}, {Volonteri}, {Haardt}, \&
  {Oh}}]{Madauetal2004}
{Madau}, P., {Rees}, M.~J., {Volonteri}, M., {Haardt}, F., \& {Oh}, S.~P. 2004,
  \apj, 604, 484

\bibitem[{{Marconi} {et~al.}(2004){Marconi}, {Risaliti}, {Gilli}, {Hunt},
  {Maiolino}, \& {Salvati}}]{Marconietal2004}
{Marconi}, A., {Risaliti}, G., {Gilli}, R., {Hunt}, L.~K., {Maiolino}, R., \&
  {Salvati}, M. 2004, \mnras, 351, 169

\bibitem[{{Marulli} {et~al.}(2006){Marulli}, {Crociani}, {Volonteri},
  {Branchini}, \& {Moscardini}}]{Marulli2006}
{Marulli}, F., {Crociani}, D., {Volonteri}, M., {Branchini}, E., \&
  {Moscardini}, L. 2006, ArXiv Astrophysics e-prints

\bibitem[{{Mayer} {et~al.}(2006){Mayer}, {Kazantzidis}, {Madau}, {Colpi},
  {Quinn}, \& {Wadsley}}]{Mayeretal2006}
{Mayer}, L., {Kazantzidis}, S., {Madau}, P., {Colpi}, M., {Quinn}, T., \&
  {Wadsley}, J. 2006, ArXiv Astrophysics e-prints

\bibitem[{{Merloni}(2004)}]{Merloni2004}
{Merloni}, A. 2004, \mnras, 353, 1035

\bibitem[{{Merloni} {et~al.}(2004){Merloni}, {Rudnick}, \& {Di
  Matteo}}]{Merlonietal2004}
{Merloni}, A., {Rudnick}, G., \& {Di Matteo}, T. 2004, \mnras, 354, L37

\bibitem[{{Merritt} {et~al.}(2004){Merritt}, {Milosavljevi{\'c}}, {Favata},
  {Hughes}, \& {Holz}}]{Merrittetal2004}
{Merritt}, D., {Milosavljevi{\'c}}, M., {Favata}, M., {Hughes}, S.~A., \&
  {Holz}, D.~E. 2004, \apjl, 607, L9

\bibitem[{{Mihos} \& {Hernquist}(1994)}]{MihosHernquist1994}
{Mihos}, J.~C. \& {Hernquist}, L. 1994, \apjl, 431, L9

\bibitem[{{Mihos} \& {Hernquist}(1996)}]{MihosHernquist1996}
---. 1996, \apj, 464, 641

\bibitem[{{Nakamura} \& {Umemura}(2001)}]{NakamuraUmemura2001}
{Nakamura}, F. \& {Umemura}, M. 2001, \apj, 548, 19

\bibitem[{{Navarro} {et~al.}(1997){Navarro}, {Frenk}, \& {White}}]{NFW1997}
{Navarro}, J.~F., {Frenk}, C.~S., \& {White}, S.~D.~M. 1997, \apj, 490, 493

\bibitem[{{Oh} \& {Haiman}(2002)}]{OhHaiman2002}
{Oh}, S.~P. \& {Haiman}, Z. 2002, \apj, 569, 558

\bibitem[{{Robertson} {et~al.}(2005){Robertson}, {Hernquist}, {Cox}, {Di
  Matteo}, {Hopkins}, {Martini}, \& {Springel}}]{Robertson2005}
{Robertson}, B., {Hernquist}, L., {Cox}, T.~J., {Di Matteo}, T., {Hopkins},
  P.~F., {Martini}, P., \& {Springel}, V. 2005, ArXiv Astrophysics e-prints

\bibitem[{{Sesana} {et~al.}(2005){Sesana}, {Haardt}, {Madau}, \&
  {Volonteri}}]{Sesanaetal2005}
{Sesana}, A., {Haardt}, F., {Madau}, P., \& {Volonteri}, M. 2005, \apj, 623, 23

\bibitem[{{Shapiro}(2005)}]{Shapiro2005}
{Shapiro}, S.~L. 2005, \apj, 620, 59

\bibitem[{{Tremaine} {et~al.}(2002){Tremaine}, {Gebhardt}, {Bender}, {Bower},
  {Dressler}, {Faber}, {Filippenko}, {Green}, {Grillmair}, {Ho}, {Kormendy},
  {Lauer}, {Magorrian}, {Pinkney}, \& {Richstone}}]{Tremaineetal2002}
{Tremaine}, S., {Gebhardt}, K., {Bender}, R., {Bower}, G., {Dressler}, A.,
  {Faber}, S.~M., {Filippenko}, A.~V., {Green}, R., {Grillmair}, C., {Ho},
  L.~C., {Kormendy}, J., {Lauer}, T.~R., {Magorrian}, J., {Pinkney}, J., \&
  {Richstone}, D. 2002, \apj, 574, 740

\bibitem[{{Volonteri} {et~al.}(2003){Volonteri}, {Haardt}, \& {Madau}}]{VHM}
{Volonteri}, M., {Haardt}, F., \& {Madau}, P. 2003, \apj, 582, 559

\bibitem[{{Volonteri} {et~al.}(2005){Volonteri}, {Madau}, {Quataert}, \&
  {Rees}}]{Volonterietal2005}
{Volonteri}, M., {Madau}, P., {Quataert}, E., \& {Rees}, M.~J. 2005, \apj, 620,
  69

\bibitem[{{Volonteri} \& {Perna}(2005)}]{VolonteriPerna2005}
{Volonteri}, M. \& {Perna}, R. 2005, \mnras, 358, 913

\bibitem[{{Volonteri} \& {Rees}(2005)}]{VolonteriRees2005}
{Volonteri}, M. \& {Rees}, M.~J. 2005, \apj, 633, 624

\bibitem[{{Walter} {et~al.}(2004){Walter}, {Carilli}, {Bertoldi}, {Menten},
  {Cox}, {Lo}, {Fan}, \& {Strauss}}]{Walteretal2004}
{Walter}, F., {Carilli}, C., {Bertoldi}, F., {Menten}, K., {Cox}, P., {Lo},
  K.~Y., {Fan}, X., \& {Strauss}, M.~A. 2004, \apjl, 615, L17

\bibitem[{{Willott} {et~al.}(2005){Willott}, {Delfosse}, {Forveille},
  {Delorme}, \& {Gwyn}}]{Willottetal2005}
{Willott}, C.~J., {Delfosse}, X., {Forveille}, T., {Delorme}, P., \& {Gwyn},
  S.~D.~J. 2005, \apj, 633, 630

\bibitem[{{Willott} {et~al.}(2003){Willott}, {McLure}, \&
  {Jarvis}}]{Willottetal2003}
{Willott}, C.~J., {McLure}, R.~J., \& {Jarvis}, M.~J. 2003, \apjl, 587, L15

\bibitem[{{Yu} \& {Tremaine}(2002)}]{YuTremaine2002}
{Yu}, Q. \& {Tremaine}, S. 2002, \mnras, 335, 965

\end{thebibliography}
\end{document}